\documentclass[aps,pra,showpacs,twocolumn]{revtex4}
\usepackage{amsmath}
\usepackage{graphicx}
\usepackage[ansinew]{inputenc}
\usepackage{array}
\usepackage{color}
\usepackage{amsxtra}
\usepackage{amstext}
\usepackage{amssymb}
\usepackage{latexsym}
\usepackage{dsfont}
\usepackage{bbold}
\usepackage{isomath}
\usepackage{pzccal}
\begin{document}

\title{Coherent cancellation of geometric phase for the OH molecule in external fields}

\author{M. Bhattacharya and S. Marin}
\affiliation{School of Physics and Astronomy, Rochester Institute of Technology, 84 Lomb Memorial Drive,
Rochester, NY 14623, USA}

\author{M. Kleinert}
\affiliation{Department of Physics, Willamette University, 900 State Street, Salem, OR 97301, USA}

\date{\today}
\begin{abstract}
The OH molecule in its ground state presents a versatile platform for precision measurement and quantum information
processing. These applications depend vitally on the accurate measurement of transition energies between the OH levels.
Significant sources of systematic errors in these measurements are shifts based on the geometric phase arising from the
magnetic and electric fields used for manipulating OH. In this article, we present these geometric phases for fields that
vary harmonically in time, as in the Ramsey technique. Our calculation of the phases is exact within the description
provided by our recent analytic solution of an effective Stark-Zeeman Hamiltonian for the OH ground state. This Hamiltonian
has earlier been shown to model experimental data accurately. We find that the OH geometric phases exhibit rich structure
as a function of the field rotation rate. Remarkably, we find rotation rates where the geometric phase accumulated by a
specific state is zero, or where the relative geometric phase between two states vanishes. We expect these findings to
be of importance to precision experiments on OH involving time-varying fields. More specifically, our analysis quantitatively
characterizes an important item in the error budget for precision spectroscopy of ground state OH.
\end{abstract}

\maketitle
\section{Introduction}
Historically, molecules were among the first physical systems to be associated with the concept of
geometric phase \cite{Mead1992,Resta2000}. In present-day molecular physics research, the phenomenon of
geometric phase plays a role of practical importance, and needs to be studied quantitatively. For example,
geometric phases need to be accurately accounted for in EDM (electron dipole moment) searches
\cite{Pendlebury2004,Lamoreaux2005,ShaferRay2008,Meyer2009,DeMille2009} and have been used to extract magnetic $g$
factors from experimental data \cite{Loh2013}. The specific effect of geometric phase we
wish to study in this article relates to precision spectroscopy of molecules.
A molecule occupying a specific internal state accumulates a dynamical phase over time. However, if the molecule
couples to a time-varying field, or has non-negligible center-of-mass motion, an additional, geometric, phase also
appears in its wavefunction. Such additional phases present themselves as energetic shifts when the molecule level
structure is probed by spectroscopic techniques such as the Ramsey method of separate oscillatory fields
\cite{Ramsey1949,Meyer2009}.

In this article we consider the $X^{2}\mathrm{\Pi}_{3/2}$ ground state manifold of the OH molecule, which contains
eight energy levels \cite{Ticknor2005,Avdeenkov2003,Sawyer2008,Tscherbul2010}. This ground manifold is presently employed in
precision measurements \cite{Hudson2006,Kozlov2009} and quantum computation \cite{Lev2006}. These experiments require
a quantitative characterization of the geometric phases \cite{Bochinski2004,Meerakker2005,Sawyer2007,Stuhl2012}.
In the presence of external electromagnetic fields, the OH ground manifold can be accurately described by an effective
eight-dimensional Hamiltonian that includes Stark and Zeeman effects \cite{Lara2008,Stuhl2012,Bohn2013}. The predictions of
this Hamiltonian have earlier been shown to agree well with experimental data \cite{Sawyer2007,Stuhl2012,Quemener2012}.
Recently, our group has presented an analytic solution to this Hamiltonian \cite{Bhattacharya2013}.

In this paper we
combine our analytic solutions with the dressed-molecule \cite{Meyer2009} as well as perturbation
\cite{DeMille2009} approaches to analyzing geometric phases. Within the
description provided by the effective Hamiltonian our results are exact, and could serve as a basis for the perturbative
inclusion of effects neglected in the model, such as hyperfine structure. It may be emphasized that our exact calculations
are valid for all rates of rotation including the case where the quanta of the rotating field are
resonant with the OH energy transitions. We use our results to discuss the lowest order, or adiabatic ``Berry" contribution
to the geometric phase, as well as higher order non-adiabatic corrections. Interestingly, we find that the geometric phase,
either belonging to a single OH state or measured relatively between two states, can often be made to vanish by choosing
the rate of field rotation appropriately. We expect this effect to be useful to the practical spectroscopy and quantum
information applications of OH.

The remainder of this paper is arranged as follows. Section ~\ref{sec:DressedApproach} summarizes the approach of Meyer \textit{et. al}
used to calculate the geometric phases in this article. Section ~\ref{sec:OHinEAndBFields} describes the Hamiltonian of the
ground state OH molecule in the presence of magnetic and electric fields. Section ~\ref{sec:OHGeometricPhaseInBField}
discusses the geometric phase for OH in a magnetic field, Section ~\ref{sec:OHGeometricPhaseInEField} in an electric field,
and Section ~\ref{sec:OHGeometricPhaseInBAndEFields} in combined magnetic and electric fields. A conclusion is supplied
at the end.

\section{Dressed molecule approach to the geometric phase}
\label{sec:DressedApproach}
We begin by summarizing the approach of Meyer \textit{et. al} to the problem of determination of geometric phases in
structured atoms and molecules \cite{Meyer2009}. In the present paper, this is the main method used for calculating
the geometric phases. The basic idea is to consider a finite-dimensional matrix Hamiltonian $H(t)$ that describes a
molecule in the presence of time-varying fields
\begin{equation}
\label{eq:TDH}
H(t)=\left(
  \begin{array}{ccc}
    h_{11}(t) & h_{12}(t) & \ldots \\
    h_{21}(t) & h_{22}(t) & \ldots \\
    \ldots & \ldots & \ldots \\
  \end{array}
\right),
\end{equation}
where $h_{ij}(t)$ where $i,j=1,2,3, \ldots$ are the generally time-dependent matrix elements. The Hamiltonian $H(t)$ satisfies the
time-dependent Schrodinger equation
\begin{equation}
\label{eq:TDSE}
i\hbar \frac{\partial \psi(t)}{\partial t}=H(t)\psi(t).
\end{equation}
In the dressed molecule approach, the wave function $\psi(t)$ is assumed to be of the form
\begin{equation}
\label{eq:WF}
\psi(t)=\left(
          \begin{array}{c}
            \alpha_{1}(t)e^{-i\omega_{1}t} \\
            \alpha_{2}(t)e^{-i\omega_{2}t} \\
            \ldots \\
          \end{array}
        \right),
\end{equation}
with $\omega_{i}, i=1,2,3,\ldots$ being parameters to be specified below. Using
Eqs.~(\ref{eq:TDH}) and ~(\ref{eq:WF}) in Eq.~(\ref{eq:TDSE}), the time-dependent Schrodinger
equation can be written as
\begin{equation}
i\hbar \frac{\partial \psi'(t)}{\partial t}=H'(t)\psi'(t),
\end{equation}
where
\begin{equation}
\psi'(t)=\left(
          \begin{array}{c}
            \alpha_{1}(t) \\
            \alpha_{2}(t) \\
            \ldots \\
          \end{array}
        \right),
\end{equation}
and
\begin{equation}
\label{eq:HamDressed}
H'(t)=W^{-1}(t)H(t)W(t)-D.
\end{equation}
The new matrices are
\begin{equation}
W(t)= \left(
  \begin{array}{ccc}
    e^{-i\omega_{1}t} & 0 & \ldots \\
    0 & e^{-i\omega_{2}t} & \ldots \\
    \ldots & \ldots & \ldots \\
  \end{array}
\right),
\end{equation}
and
\begin{equation}
D = \hbar\left(
  \begin{array}{ccc}
    \omega_{1} & 0 & \ldots \\
    0 & \omega_{2} & \ldots \\
    \ldots & \ldots & \ldots \\
  \end{array}
\right).
\end{equation}
With an appropriate selection of the parameters $\omega_{i}$, the matrix $H'(t)$ can be made independent of time.
The resulting matrix will be called $H_{d}$, the dressed matrix. The ``dressed" eigenvalues of $H_{d}$ include
the ``bare" molecular energies as well as the energy contributions due to the time-varying fields. In this article,
we will only consider fields varying harmonically with time at a rotation frequency $\omega_{r}$
\cite{Ramsey1949,Meyer2009,DeMille2009,Loh2013},
implying therefore that the corresponding quanta carry an energy $\hbar\omega_{r}$. The difference between the
dressed and bare energies will be used to find the geometric phase, as specified below.

\section{Ground state OH Stark-Zeeman Hamiltonian}
\label{sec:OHinEAndBFields}
In this section we present the matrix Stark-Zeeman Hamiltonian that describes OH in the $X^{2}\mathrm{\Pi}_{3/2}$ state
\cite{Stuhl2012,Lara2008}
\begin{equation}
\label{eq:H1}
H_{m}=H_{0}-\vec{\mu}_{e}\cdot \vec{E}-\vec{\mu}_{b}\cdot \vec{B},
\end{equation}
where $H_{0}$ describes the internal structure of the molecule.  The second and third terms correspond to the
interaction of the OH electric ($\vec{\mu}_{e}$) and magnetic ($\vec{\mu}_{b}$) dipole moments with externally
applied electric ($\vec{E})$ and magnetic $(\vec{B}$) fields, respectively. This model is accurate for electric
fields stronger than 1 kV/cm and magnetic fields above 100 G \cite{Ticknor2005}, or for OH vapor temperatures higher
than 5 mK. In all these cases, hyperfine structure can be neglected. This model Hamiltonian has shown very good
agreement with data from OH experiments on nonadiabatic transitions \cite{Stuhl2012} and evaporative
cooling \cite{Quemener2012}. Thus, our use of the Hamiltonian of Eq.~(\ref{eq:H1}) for calculating geometric phases
in OH is justified theoretically as well as experimentally.

In the OH ground state manifold, Eq.~(\ref{eq:H1}) can be represented as an eight dimensional matrix using the Hund's
case (a) parity basis $|J,M,\bar{\Omega},\epsilon \rangle$ \cite{Lara2008}. Here
$J=3/2$ is the total molecular angular momentum, $M$ its laboratory frame projection, $\bar{\Omega}$ its component
along the internuclear axis, and $\epsilon=\{e,f\}$ is the $e-f$ symmetry. Using the matrix elements provided in
the appendix of Ref.~\cite{Lara2008} we can write the eight-dimensional matrix representation as
\begin{equation}
\label{eq:HamUndressed}
H_{M}=\left(
        \begin{array}{cc}
          P & Q \\
          Q^{\dagger} & R \\
        \end{array}
      \right).
\end{equation}
The $4\times4$ matrices $P,Q$ and $R$ are
\begin{widetext}

\begin{align}
\begin{split}
\label{eq:PQR}
P=&\left(
  \begin{array}{cccc}
     -\frac{1}{2}\hbar\Delta -\frac{6}{5} \mu _{B} B_{z}  & \frac{2\sqrt{3}}{5}  \mu _{B} \left(B_{x}+i B_{y}\right)  & 0 & 0 \\
     \frac{2\sqrt{3}}{5} \mu _{B} \left(B_{x}-i B_{y}\right)  & -\frac{1}{2}\hbar \Delta-\frac{2}{5}\mu _{B} B_{z}  &
     \frac{4}{5}\mu _{B} \left(B_{x}+i B_{y}\right)  & 0 \\
    0 & \frac{4}{5}  \mu _{B} \left(B_{x}-i B_{y}\right) & -\frac{1}{2}\hbar\Delta+ \frac{2}{5}\mu _{B}B_{z} & \frac{2\sqrt{3}}{5} \mu _{B}\left(B_{x}+i B_{y}\right)\\
    0 & 0 & \frac{2\sqrt{3}}{5}\mu _{B}\left(B_{x}-i B_{y}\right) & -\frac{1}{2}\hbar\Delta+\frac{6}{5}\mu _{B}B_{z} \\
  \end{array}
\right),\\
Q=&\left(
    \begin{array}{cccc}
      \frac{3}{5}\mu _{e}E_{z} &
   -\frac{\sqrt{3}}{5}\mu _{e}\left(E_{x}+iE_{y}\right) & 0 & 0 \\
      -\frac{\sqrt{3}}{5}\mu _{e}\left(E_{x}-iE_{y}\right) & \frac{1}{5}\mu _{e}E_{z} & -\frac{2}{5}\mu _{e}\left(E_{x}+i
       E_{y}\right)  & 0  \\
      0 & -\frac{2}{5}\mu _{e}\left(E_{x}-iE_{y}\right)& -\frac{1}{5}\mu _{e}E_{z} & -\frac{\sqrt{3}}{5}
    \mu _{e}\left(E_{x}+i E_{y}\right) \\
      0 & 0 & -\frac{\sqrt{3}}{5}\mu _{e}\left(E_{x}-i E_{y}\right) & -\frac{3}{5}\mu _{e}E_{z}  \\
    \end{array}
  \right),\\
R =& P + \hbar \Delta I_4.
\end{split}
\end{align}

\end{widetext}
where $\Delta$ is the lambda-doubling constant, $\mu_{B}$ is the Bohr magneton, $\mu_{e}$ the molecular electric
dipole moment, $I_4$ is the four-dimensional identity matrix, and $E_{x},E_{y}$ and $E_{z}(B_{x},B_{y}$ and $B_{z})$ are the Cartesian components of the electric
(magnetic) fields, respectively.

\section{Geometric phase of OH in a magnetic field}
\label{sec:OHGeometricPhaseInBField}
\begin{figure}[h!]
\label{fig:F1}
\includegraphics[width=0.99\columnwidth]{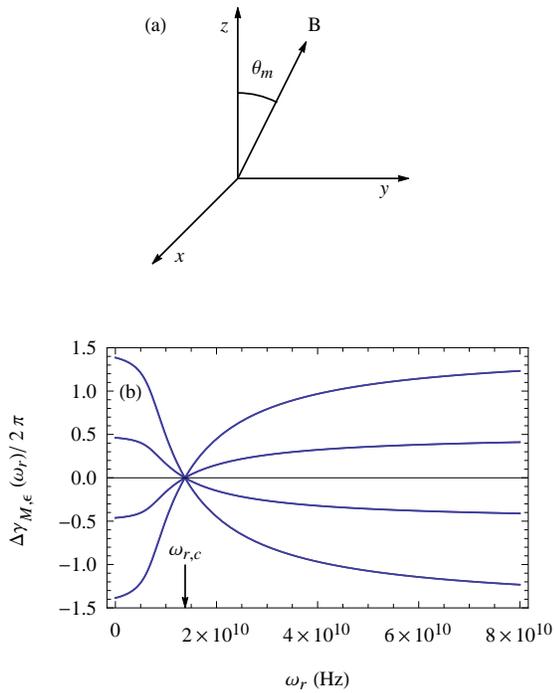}
\caption{(a) The magnetic field rotating about the laboratory $z$ axis while making an angle $\theta_{m}$. (b) The
geometric phase for the OH state $\gamma_{M,\epsilon}$ [Eq.~(\ref{eq:LabProjBerry})] as a function of the magnetic
field rotation frequency $\omega_{r}$. The parameters are $\theta_{m}=\pi/8$ and $B=0.1$ T. The
critical rotation frequency $\omega_{r,c} \simeq$ \mbox{13.8 GHz} [Eq.~(\ref{eq:CritRot})], where all the geometric phases
are equal (to zero), is indicated with an arrow.}
\label{fig:P1}
\end{figure}

First we consider the ground state OH molecule in a magnetic field rotating at a frequency $\omega_{r}$ at an angle
$\theta_{m}$ with respect to the laboratory $z$ axis, see Fig.~\ref{fig:P1}(a). This situation is
relevant to pure magnetic trapping of OH \cite{Sawyer2008}. We use
\begin{eqnarray}
B_{x}=B\sin\theta_{m}\cos\omega_{r}t,& B_{y}=B\sin\theta_{m}\sin\omega_{r}t,\nonumber\\
B_{z}=B\cos\theta_{m},& E_{x}=E_{y}=E_{z}=0,\\
\nonumber
\end{eqnarray}
in Eq.~(\ref{eq:PQR}) and plug in the resulting Hamiltonian of Eq.~(\ref{eq:HamUndressed}) into
Eq.~(\ref{eq:HamDressed}). Subsequently, we find, that with the choice of parameters
\begin{eqnarray}
\label{eq:FreqParam}
\omega_{1}=\omega_{5}=-\omega_{4}=-\omega_{8}=-3\omega_{r}/2,&\nonumber\\
\omega_{2}=\omega_{6}=-\omega_{3}=-\omega_{7}=-\omega_{r}/2,&\\
\nonumber
\end{eqnarray}
the Hamiltonian $H'$ can be made time-independent, yielding the dressed matrix
\begin{equation}
\label{eq:HamOHBPrime}
H_{d}=\left(
        \begin{array}{cc}
          P' & 0 \\
          0 & R' \\
        \end{array}
      \right),
\end{equation}
where
\small
\begin{widetext}
\begin{align}
\begin{split}
\label{eq:PQRPrimeOHB}
P'=&\left(
  \begin{array}{cccc}
     \frac{3}{2}\hbar\omega_{r}-\frac{1}{2}\hbar\Delta -\frac{6}{5} \mu _{B}B \cos\theta_{m}  & \frac{2\sqrt{3}}{5}  \mu _{B} B\sin\theta_{m} & 0 & 0 \\
     \frac{2\sqrt{3}}{5} \mu _{B} B \sin\theta_{m}  & \frac{1}{2}\hbar\omega_{r}-\frac{1}{2}\hbar \Delta-\frac{2}{5}\mu _{B} B \cos\theta_{m}  &
     \frac{4}{5}\mu _{B} B\sin\theta_{m}  & 0 \\
    0 & \frac{4}{5}  \mu _{B} B\sin\theta_{m} & -\frac{1}{2}\hbar\omega_{r}-\frac{1}{2}\hbar\Delta+ \frac{2}{5}\mu _{B}B\cos\theta_{m} & \frac{2\sqrt{3}}{5} \mu _{B}B\sin\theta_{m}\\
    0 & 0 & \frac{2\sqrt{3}}{5}\mu _{B}B\sin\theta_{m} & -\frac{3}{2}\hbar\omega_{r}-\frac{1}{2}\hbar\Delta+\frac{6}{5}\mu _{B}B \cos\theta_{m} \\
  \end{array}
\right),\\
R'=& P' + \hbar \Delta I_4.
\end{split}
\end{align}

\normalsize
The eight eigenvalues of the dressed matrix are
\begin{equation}
\tilde{E}_{M,\epsilon}(\omega_{r})=\frac{\epsilon}{2}\hbar\Delta
+M \hbar \sqrt{\omega_L^{2}+ \omega_{r}^{2}-2\omega_L\omega_{r}\cos\theta_{m}},
\end{equation}
\end{widetext}
where $M=-3/2,-1/2,1/2,3/2$, $\epsilon=(-1,1)$ marks the $e-f$ parity, and \begin{equation}
\label{eq:LarmorMagnetic}
\omega_{L} = \frac{4}{5} \frac{\mu_{B}B}{ \hbar},
\end{equation}
is the Larmor frequency. The total phase picked up by an OH state in
a time $2\pi/\omega_{r}$ is given by
\begin{equation}
\gamma_{M,\epsilon}(\omega_{r})=\frac{\tilde{E}_{M,\epsilon}(\omega_{r})}{\hbar}\frac{2\pi}{\omega_{r}},
\end{equation}
and the total geometric phase is found by subtracting the dynamical phase from the total phase,
\begin{equation}
\Delta\gamma_{M,\epsilon}(\omega_{r})=\gamma_{M,\epsilon}(\omega_{r})-\gamma_{M,\epsilon}(0).
\end{equation}
The geometric phase scaled by the laboratory projection of the angular momentum $M$ for each state is
\begin{equation}
\label{eq:LabProjBerry}
\frac{\Delta\gamma_{M,\epsilon}(\omega_{r})}{M}=
\frac{2\pi}{\omega_{r}}\left(\sqrt{\omega_{L}^{2}+\omega_{r}^{2}-2\omega_{L}\omega_{r}\cos\theta_{m}}-\omega_{L}\right).
\end{equation}
 Note that the geometric phase is independent of $\Delta$ and also of the parity $\epsilon$.
In Fig.~\ref{fig:P1}(b)
curves for various $\theta_{m}$ are shown as a function of the field rotation rate $\omega_{r}$. The critical value of
rotation at which all the states have vanishing geometric phase can be found readily from Eq.~(\ref{eq:LabProjBerry}),
\begin{equation}
\label{eq:CritRot}
\omega_{r,c}=2\omega_{L}\cos \theta_{m}.
\end{equation}
At this value of rotation, the adiabatic contribution to the geometric phase is canceled exactly by the non-adiabatic
phase. Presumably this is a convenient rate for performing spectroscopy, as well as useful for quantum information
processing \cite{Leek2007}. In the adiabatic case, $\omega_{r}/\omega_{L} \ll 1$ and the lowest order or Berry phase is
\begin{equation}
\label{eq:BerryResultMagnetic}
\frac{\Delta\gamma_{M,\epsilon}^{(0)}(\omega_{r})}{M} \simeq -2\pi \cos \theta_{m},
\end{equation}
which can be related to the standard result (see Ref.~\cite{BohmBook} and references therein)
\begin{equation}
\label{eq:BerryStandard}
\frac{\Delta\gamma_{M,\epsilon}^{(0)}(\omega_{r})}{M} = 2\pi (1-\cos \theta_{m}),
\end{equation}
by adding our result of Eq.~(\ref{eq:BerryResultMagnetic}) to $2\pi.$ This manipulation does not change
the relative phases between the states, to which  Ramsey-type experiments are sensitive \cite{Meyer2009}. Higher
order non-adiabatic corrections can readily be obtained from Eq.~(\ref{eq:LabProjBerry}). For example, the next
order term in $\omega_{r}/\omega_{L}$ is
\begin{equation}
\label{eq:FirstNonAdia}
\frac{\Delta\gamma_{M,\epsilon}^{(1)}(\omega_{r})}{M} = 2\pi \sin^{2} \theta_{m}\left(\frac{\omega_{r}}{2\omega_{L}}\right).
\end{equation}
Also, the limit of high rotation can be considered, where $\omega_{m} \gg \omega_{L}.$ In this case,
\begin{equation}
\frac{\Delta\gamma_{M,\epsilon}(\omega_{r})}{M} \simeq 2\pi-2\pi (1+\cos \theta_{m})\left(\frac{\omega_{L}}{\omega_{r}}\right).
\end{equation}
In the limit of very fast rotation, the states accumulate a phase of $2\pi M$ [see Fig.~\ref{fig:P1}(b)].

\section{Geometric phase of OH in an electric field}
\label{sec:OHGeometricPhaseInEField}
\begin{figure}
\includegraphics[width=0.95\columnwidth]{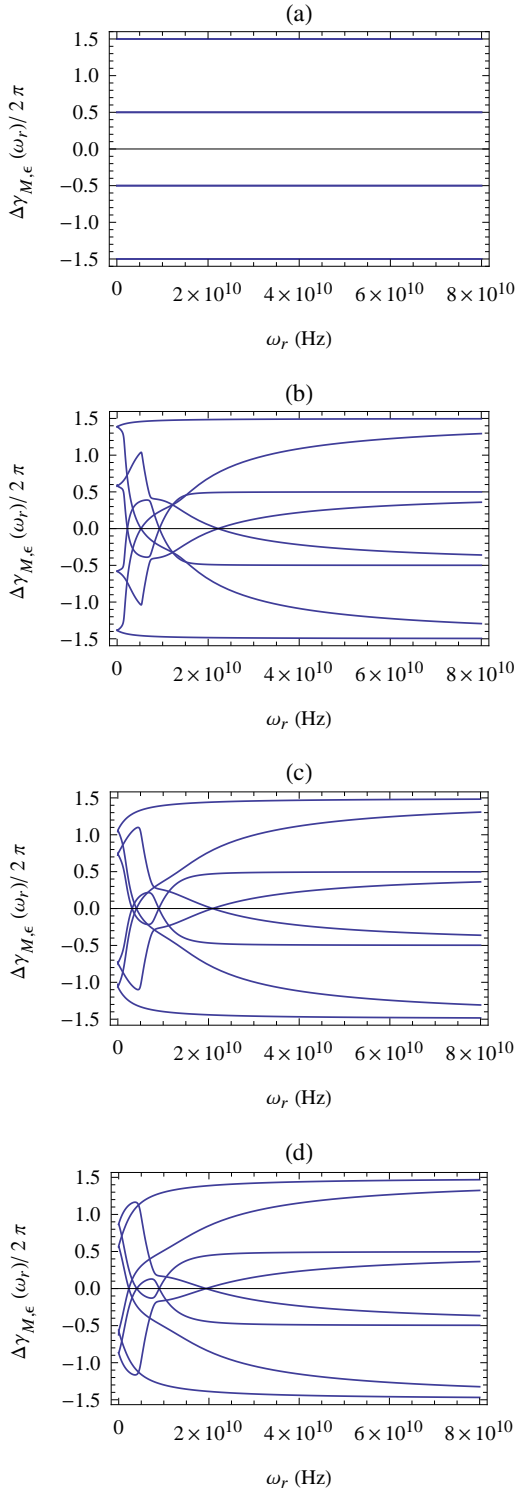}
\caption{The geometric phase for the OH state with angular momentum projection $M$ [Eq.~(\ref{eq:LabProjBerry})] as a
function of the electric field rotation frequency $\omega_{r}$. The parameters are
$\Delta=1.66$GHz, $\mu_{e}=1.667$ D, $E=2$ kV/cm, and (a) $\theta_{e}=0,$ (b) $\theta_{e}=\pi/8$, (c) $\theta_{e}=\pi/4$
and (d) $\theta_{e}=3\pi/8$.}
\label{fig:P2}
\end{figure}

In this section we consider the ground state OH molecule in an electric field rotating at a frequency $\omega_{r}$ at an
angle $\theta_{e}$ with respect to the laboratory $z$ axis. This situation may be relevant to electrostatic
slowing and trapping of OH \cite{Meerakker2005}. For simplicity and in preparation for section \ref{sec:OHGeometricPhaseInBAndEFields},
we have chosen the same rotation frequency $\omega_{r}$
for the electric field as for the magnetic field above, but a different angle of inclination to the $z$ axis. We use
\begin{eqnarray}
B_{x}=B_{y}=B_{z}=0,&&\\
E_{x}=E\sin\theta_{e}\cos\omega_{r}t,& E_{y}=E\sin\theta_{e}\sin\omega_{r}t,& E_{z}=E\cos\theta_{e},\nonumber\\
\nonumber
\end{eqnarray}
in Eq.~(\ref{eq:PQR}) and plug in the resulting Hamiltonian of Eq.~(\ref{eq:HamUndressed}) into
Eq.~(\ref{eq:HamDressed}). Subsequently, we find, that with the choice of parameters of Eq.~(\ref{eq:FreqParam})
the Hamiltonian $H'$ can be made time-independent, yielding the dressed matrix
\begin{equation}
\label{eq:HamOHEPrime}
H_{d}=\left(
        \begin{array}{cc}
          P' & Q' \\
          Q' & R' \\
        \end{array}
      \right),
\end{equation}
where
\small
\begin{widetext}
\begin{align}
\label{eq:PQRPrimeOHB}
\begin{split}
P'=&\left(
  \begin{array}{cccc}
     \frac{3}{2}\hbar\omega_{r}-\frac{1}{2}\hbar\Delta  & 0 & 0 & 0 \\
     0  & \frac{1}{2}\hbar\omega_{r}-\frac{1}{2}\hbar \Delta  &
     0  & 0 \\
    0 & 0 &- \frac{1}{2}\hbar\omega_{r}-\frac{1}{2}\hbar\Delta & 0\\
    0 & 0 & 0 & -\frac{3}{2}\hbar\omega_{r}-\frac{1}{2}\hbar\Delta\\
  \end{array}
\right),\\
Q'=&\left(
    \begin{array}{cccc}
      \frac{3}{5}\mu _{e}E\cos\theta_{e} &
   -\frac{\sqrt{3}}{5}\mu _{e}E\sin\theta_{e} & 0 & 0 \\
      -\frac{\sqrt{3}}{5}\mu _{e}E\sin\theta_{e} & \frac{1}{5}\mu _{e}E\cos\theta_{e} & -\frac{2}{5}\mu _{e}E\sin\theta_{e}  & 0  \\
      0 & -\frac{2}{5}\mu _{e}E\sin\theta_{e}& -\frac{1}{5}\mu _{e}E\cos\theta_{e} & -\frac{\sqrt{3}}{5}
    \mu _{e}E\sin\theta_{e} \\
      0 & 0 & -\frac{\sqrt{3}}{5}\mu _{e}E\sin\theta_{e} & -\frac{3}{5}\mu _{e}E\cos\theta_{e}  \\
    \end{array}
  \right),\\
R'=& P' + \hbar \Delta I_4.
\end{split}
\end{align}
\end{widetext}
\normalsize
The eigenvalues of the corresponding dressed matrix $H_{d}$ were found analytically using the techniques of
Ref.~\cite{Bhattacharya2013}. They are provided in the Supplementary Material \cite{Supp}. The corresponding geometric
phases are plotted in Fig.~\ref{fig:P2}. It can be seen that unlike Fig.~\ref{fig:P1}(b), the variation of
geometric phase with electric field rotation rate exhibits a rich structure, including crossings and avoided
crossings, and offering possibilities for coherent control of the OH geometric phase \cite{Leek2007}. We will now
make several comments about these geometric phase plots.

The (avoided) crossings in Fig.~\ref{fig:P2} may be traced back to similar structures in
the dressed eigenvalue spectrum. We note that while the geometric phases in Fig.~\ref{fig:P2} exhibit rather strong
kinks and turns, the dressed energies behave much more regularly and display (avoided) crossings when two states
approach one another. The presence of the (avoided) crossings may be explained by standard Floquet theory \cite{Grifoni1998}.
Specifically, since the Hamiltonian $H_{M}$ is periodic in time,
\begin{equation}
H_{M}\left(t+\frac{2\pi}{\omega_{r}}\right)=H_{M}(t),
\end{equation}
the dressed states are either odd or even under the same symmetry. We emphasize that in the presence of all three
components of the electric field neither parity nor any component of the angular momentum are conserved. As
the parameter $\omega_{r}$ is varied, states with the same symmetry avoid each other in the spectrum, while
those with unlike symmetry are allowed to cross, according to the Wigner von-Neumann theorem \cite{Neumann1929}.
The structure of the eigenvalues is so complicated that simple analytical expressions for the rotation rates at
which the (avoided) crossings occur are not easy to derive.

The crossings in Fig.~\ref{fig:P2}, unlike those in Fig.~\ref{fig:P1}, do not all occur at the same rotation rate.
Nonetheless, they offer similar possibilities for implementing geometric-phase-free spectroscopy and quantum
information storage \cite{Leek2007}.  As there are no incoherent
mechanisms present in the OH model, it may be said that the (non-zero) zero-phase crossings in Fig.~\ref{fig:P2}
as well as Fig.~\ref{fig:P1} correspond to the \textit{coherent} cancellation of (relative) geometric phase. The
cancellation occurs due to interferometric destruction of multiple pathways for accumulating geometric phase.
This phenomenon seems analogous to the coherent suppression of tunneling described earlier for driven quantum
systems \cite{Grifoni1998}.

It should be noticed that the geometric phase plots possess a reflection symmetry about the horizontal,
zero-phase, axis. It is this chiral symmetry which allows for the analytic determination of the phases, in a
manner similar to that shown originally for the eigenvalues of the time-independent OH spectrum \cite{Bhattacharya2013}.

Lastly, we note that for very fast rotation the OH states accumulate phases $2\pi M$, as expected.
\begin{figure}
\includegraphics[width=0.95\columnwidth]{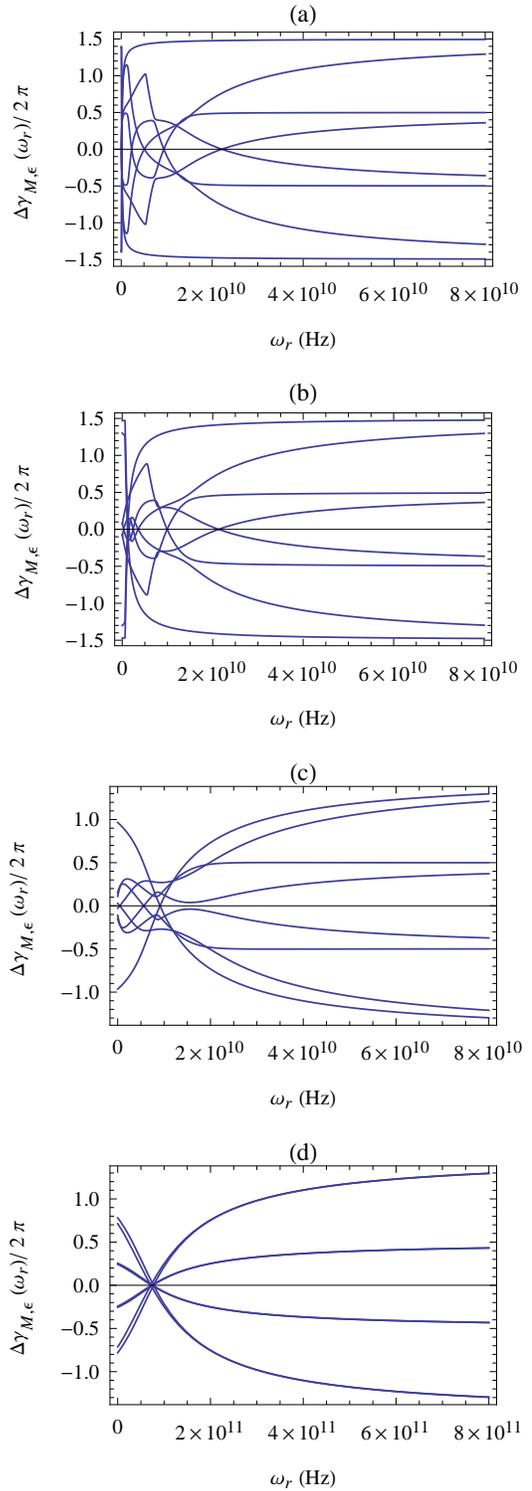}
\caption{The geometric phase for the OH state with angular momentum projection $M$ [Eq.~(\ref{eq:LabProjBerry})] as
a function of the electric and magnetic field rotation frequency $\omega_{r}$. The
parameters used are $\Delta=1.66$ GHz, $\mu_{e}=1.667$ D, $E=2$ kV/cm, $\theta_{e}=\pi/8,$ $\theta_{m}=\pi/3$ and
(a) $B=0.001$ T (b) $B=0.01$ T, (c) $B=0.1$ T and (d) $B=1.0$ T.}
\label{fig:P3}
\end{figure}

From the exact solutions, the expressions for the low order phases can easily be obtained. For convenience we divide
our discussion of the approximate results into two parts below, which refer to weak and strong electric fields, respectively.

\subsection{Weak electric fields: $\mu_{e}E \ll \hbar\Delta$}
Here we consider the situation where the Stark shifts are smaller than the lambda-doublet splitting. Without loss of
generality, we present the example of the most energetic state in the manifold $(M=3/2,\epsilon=f)$. For low rotation rate
$\omega_{r}$, the Berry phase is
\begin{equation}
\label{eq:BerryResultElectric}
\Delta\gamma_{3/2,f}^{(0)}(\omega_{r}) \simeq 2\pi \left(\frac{3}{2}\right)\cos \theta_{e}.
\end{equation}
The first non-adiabatic correction is
\begin{equation}
\label{eq:NonAdiaWeakElectric}
\Delta\gamma_{3/2,f}^{(1)}(\omega_{r})
\simeq 2\pi \left(\frac{75\Delta}{32\omega_{e}^{2}}
+\frac{9}{16\Delta}+\frac{81 \omega_{e}^{2}}{400\Delta^{3}}\right)\sin^{2}\theta_{e}\omega_{r},
\end{equation}
where
\begin{equation}
\omega_{e}=\frac{\mu_{e}E}{\hbar},
\end{equation}
is the electric counterpart of the magnetic Larmor frequency of Eq.~(\ref{eq:LarmorMagnetic}).
In order for the first term in the parentheses of Eq.~(\ref{eq:NonAdiaWeakElectric}) to yield a
small correction for weak fields (i.e. small $\omega_{e}$), the rotation rate should obey the condition
\begin{equation}
\omega_{r} \ll \frac{\omega_{e}^{2}}{\Delta},
\end{equation}
which corresponds to the rotation rate being smaller than the quadratic Stark shift. It should be noted that
the presence of the $\omega_{e}$ in the denominator of the first term in Eq.~(\ref{eq:NonAdiaWeakElectric})
indicates that perturbative expansion breaks down in the limit of vanishing electric field \cite{Stuhl2012},
also see Section \ref{sec:OHGeometricPhaseInBAndEFields} below. Likewise, in order for the second term in the
parentheses to yield a small correction,
\begin{equation}
\omega_{r} \ll \Delta,
\end{equation}
and for the third term
\begin{equation}
\omega_{r} \ll \frac{\Delta^{3}}{\omega_{e}^{2}}.
\end{equation}

\subsection{Strong electric fields: $\mu_{e}E \gg \hbar\Delta$}
Now we consider the case where the electric field is large enough that the Stark shift is greater than the
lambda-doublet splitting. The Berry phase $\Delta\gamma_{3/2,f}^{(0)}(\omega_{r})$ is identical to
Eq.~(\ref{eq:BerryResultElectric}). The first nonadiabatic correction is
\begin{equation}
\label{eq:NonAdiaStrongElectric}
\Delta\gamma_{3/2,f}^{(1)}(\omega_{r})
\simeq 2\pi \left(\frac{15}{8\omega_{e}}
+\frac{125 \Delta^{2}}{96\omega_{e}^{3}}\right)\sin^{2}\theta_{e}\omega_{r},
\end{equation}
where we have retained only the lowest order term in $\Delta/\omega_{e}$. For the correction of
Eq.~(\ref{eq:NonAdiaStrongElectric}) to be small, the rotation rate requires
\begin{equation}
\omega_{r} \ll \left(\omega_{e}, \frac{\omega_{e}^{3}}{\Delta^{2}}\right).
\end{equation}
We note that for strong electric fields, the regime of linear Stark effect is reached and the nonadiabatic
corrections now depend on odd powers of the electric field. Higher order terms in the geometric phase can
be extracted from our exact solutions in a similar manner. Likewise, a more detailed examination of the
physical mechanism behind the accumulation of geometric phase by other states in the ground state OH
manifold can be carried out readily \cite{Meyer2009}.

\section{Geometric phase of OH in combined $B$ and $E$ fields}
\label{sec:OHGeometricPhaseInBAndEFields}
We now consider the geometric phases accumulated by the ground state OH molecule in the combined presence
of both magnetic and electric fields. This situation corresponds to the magnetoelectrostatic manipulation of OH
\cite{Sawyer2007,Stuhl2012,Quemener2012}. We will first present an exact solution to the problem, and then a
perturbation theory approach.

\subsection{Exact results}
For the time-varying magnetic and electric fields we assume
\begin{eqnarray}
\label{eq:SimulBAndEComponents}
B_{x}=B\sin\theta_{m}\cos\omega_{r}t,& B_{y}=B\sin\theta_{m}\sin\omega_{r}t,\nonumber\\
B_{z}=B\cos\theta_{m},&\\
E_{x}=E\sin\theta_{e}\cos\omega_{r}t,& E_{y}=E\sin\theta_{e}\sin\omega_{r}t,\nonumber\\
E_{z}=E\cos\theta_{e}.&\nonumber\\
\nonumber
\end{eqnarray}
Following the procedure described in the previous sections, a time-independent dressed matrix $H_{d}$ was found.
Using the eigenvalues of the dressed matrix supplied in the Supplementary Material \cite{Supp}, the total
geometric phase for each state has been plotted as a
function of the rotation rate $\omega_{r}$ in Fig.~\ref{fig:P3}. Although the specific structure in this case is
different from that of Fig.~\ref{fig:P2}, (avoided) crossings are still present. It can be seen that as the magnitude
of the magnetic field is increased from (a)-(d), Fig.~\ref{fig:P3} begins to resemble Fig.~\ref{fig:P1}. Thus, for
very strong magnetic fields, the electric field contribution becomes less significant, and the geometric phases can
be made almost all equal (to zero) near the critical rotation frequency.
\subsection{Time-dependent perturbation theory}
In the limit of small geometric phases, a quicker route to finding low order terms is provided by time-dependent
perturbation theory \cite{DeMille2009}. The limit of small phases corresponds in our case to assuming small $\theta_{m}$ and
$\theta_{e}$, as these determine the solid angle traversed geometrically during rotation. To implement this approach,
we use the fields of Eq.~(\ref{eq:SimulBAndEComponents}) in Eq.~(\ref{eq:PQR}) and write the resulting Hamiltonian
of Eq.~(\ref{eq:HamUndressed}) in a form suitable for Floquet perturbation theory \cite{Sambe1973},
\begin{equation}
\label{eq:HamFloquet}
H_{M}=H_{u}+V_{s}+V_{-}e^{-i\omega_{r}t}+V_{+}e^{+i\omega_{r}t},
\end{equation}
where $H_{u}$ is diagonal and describes the lambda doublet states in the presence of a time-independent magnetic
field $B_{z},$ $V_{s}$ is the nondiagonal static part of the perturbation due to the constant
electric field $E_{z}$ and $V_{-}=V_{+}^{\dagger}$ are the coefficient matrices of the time
dependent functions $e^{\mp i\omega_{r}t},$ respectively. These matrices are
\begin{widetext}
\begin{equation}
\label{eq:TDmatrices}
H_{u} =
\small
\begin{pmatrix}
  -\frac{\hbar\Delta}{2}-\frac{6}{5}\mu_{B}B_{z} & 0 &  0  & 0 & 0   & 0            &    0         & 0  \\
           0     & -\frac{\hbar\Delta}{2}-\frac{2}{5}\mu_{B}B_{z}   & 0    &      0 & 0 &0           &  0           &  0 \\
           0     & 0       &  -\frac{\hbar\Delta}{2}+\frac{2}{5}\mu_{B}B_{z}   &   0  &  0      &   0            &             0     & 0 \\
           0&0&0&-\frac{\hbar\Delta}{2}+\frac{6}{5}\mu_{B}B_{z} &0 &0 &0 &0 \\
0  & 0 &       0        & 0  & \frac{\hbar\Delta}{2}-\frac{6}{5}\mu_{B}B_{z}  &             0&     0        &  0 \\
0   & 0 & 0 & 0 & 0  & \frac{\hbar\Delta}{2}-\frac{2}{5}\mu_{B}B_{z}             &   0          &  0 \\
       0    &  0 & 0& 0 & 0 & 0   &\frac{\hbar\Delta}{2}+\frac{2}{5}\mu_{B}B_{z}   & 0 \\
       0    &  0     & 0 & 0 &   0  & 0 &0 &\frac{\hbar\Delta}{2}+\frac{6}{5}\mu_{B}B_{z}   \\
\end{pmatrix},\\
\end{equation}
\begin{equation}
V_{s} =
\small
\begin{pmatrix}
  0      &     0         &    0        &       0 & \frac{3}{5}\mu_{e}E_{z}   & 0            &    0         & 0  \\
           0     & 0    & 0    &      0 & 0  & \frac{1}{5}\mu_{e}E_{z}            &  0           &  0 \\
           0     & 0       &  0  &   0  &  0      &   0 & -\frac{1}{5}\mu_{e}E_{z}       & 0  \\
           0&0&0&0 &0 &0 &0&-\frac{3}{5}\mu_{e}E_{z} \\
\frac{3}{5}\mu_{e}E_{z}  & 0  &       0        & 0  & 0  &             0&     0        &  0 \\
0   & \frac{1}{5}\mu_{e}E_{z} & 0 & 0 & 0  & 0            &   0          &  0 \\
       0    &  0 & -\frac{1}{5}\mu_{e}E_{z}  & 0 & 0 & 0   &0  & 0 \\
       0    &  0     & 0  & -\frac{3}{5}\mu_{e}E_{z} &   0  & 0 &0 &0   \\
\end{pmatrix},\\
\end{equation}
\begin{equation}
V_{-} =i
\small
\begin{pmatrix}
 0 & \frac{2\sqrt{3}}{5}\mu_{B}B_{l}  & 0 & 0 & 0 & -\frac{\sqrt{3}}{5}\mu_{e} E_{l}  & 0 & 0 \\
 0 & 0 & \frac{4}{5}\mu_{B}B_{l} & 0 & 0 & 0 & -\frac{2}{5}\mu_{e} E_{l} & 0 \\
 0 & 0 & 0 & \frac{2\sqrt{3}}{5}\mu_{B}B_{l}  & 0 & 0 & 0 & -\frac{\sqrt{3}}{5}\mu_{e} E_{l} \\
 0 & 0 & 0 & 0 & 0 & 0 & 0 & 0 \\
 0 & -\frac{\sqrt{3}}{5}\mu_{e} E_{l} & 0 & 0 & 0 & \frac{2\sqrt{3}}{5}\mu_{B}B_{l}  & 0 & 0 \\
 0 & 0 & -\frac{2}{5}\mu_{e} E_{l} & 0 & 0 & 0 & \frac{4}{5}\mu_{B}B_{l} & 0 \\
 0 & 0 & 0 & -\frac{\sqrt{3}}{5}\mu_{e}E_{l} & 0 & 0 & 0 & \frac{2\sqrt{3}}{5} \mu_{B}B_{l} \\
 0 & 0 & 0 & 0 & 0 & 0 & 0 & 0 \\
\end{pmatrix},\\
\end{equation}
\end{widetext}
where
\begin{equation}
B_{l}=B\sin\theta_{m}, E_{l}=E\sin\theta_{e}.
\end{equation}
Assuming that the parameters are arranged to operate away from the crossings in the spectrum of the unperturbed Hamiltonian
$H_{u}$ \cite{Cawley2013}, we use non-degenerate time-dependent perturbation theory. Our assumption of small angles
$\theta_{e}$ and $\theta_{m}$ implies that the quantities $V_s$, $V_-$, and $V_+$ (from here on collectively referred to as ``$V$") are small in Eq.~(\ref{eq:HamFloquet}), and may be used as
perturbation parameters.

\subsubsection{Second order}
From the shifts to the bare eigenvalues, the geometric phases can then be found. As in the earlier sections, without loss of
generality, we demonstrate our results on the most energetic state in the ground state manifold $(J=3/2,\epsilon=f)$. From
perturbation theory, we find there are no corrections to first order in the $V$ operators of Eq.~(\ref{eq:HamFloquet}).
The lowest correction is second order in $V$, and yields for the geometric phase,
\begin{equation}
\label{eq:BerryFloquet}
\Delta\gamma_{M,\epsilon}^{0}\simeq 2\pi\left[\frac{3}{4}\tan^{2}\theta_{m}
+\frac{3\left(\omega_{e}\sin\theta_{e}\right)^{2}}{\left(5\Delta+\omega_{L}\cos\theta_{m}\right)^{2}}\right].
\end{equation}
We may compare this equation to our previous results. It can be readily seen that in the case of a zero electric field
$(\omega_{e}=0)$, and small $\theta_{m}$ both Eq.~(\ref{eq:BerryFloquet}) and Eq.~(\ref{eq:BerryStandard}) reduce to
\begin{equation}
\Delta\gamma_{M,\epsilon}^{0}\simeq 2\pi\left(\frac{3}{2}\right)\frac{\theta_{m}^{2}}{2}.
\end{equation}
Thus the results agree well in the case of a pure magnetic field.

However, in the case of a pure electric field, the result of Eq.~(\ref{eq:BerryFloquet}) does not connect smoothly with
that of Eq.~(\ref{eq:BerryResultElectric}). In fact it can be shown that the result of the perturbative approach diverges
at zero magnetic field. This is not surprising, since we assumed a lack of degeneracy in the bare spectrum so that we could
use non-degenerate perturbation theory to derive Eq.~(\ref{eq:BerryFloquet}). However, in the absence of a magnetic field,
each lambda-doublet state of the bare spectrum is in fact four-fold degenerate [as can be seen readily from
Eq.~(\ref{eq:TDmatrices})], contradicting our assumption. A correct
approach in this case would be to use time-dependent degenerate perturbation theory; we do not follow this route further.
However, we do note that this discussion underscores the importance of our exact results in the case of the OH molecule in a pure
electric field. Nonadiabatic corrections can also be found at this level of perturbation theory, but we do not discuss them here.
\subsubsection{Third order}
The next order of perturbation theory in the $V$ operators of Eq.~(\ref{eq:HamFloquet}) yields a correction to the Berry
phase
\begin{eqnarray}
\Delta\gamma_{M,\epsilon}^{0,1}&=&2\pi\sin\theta_{m}\sin 2\theta_{e}\left(\frac{2\omega_{e}^{2}}{5\Delta\omega_{L}}\right)\\
&\times &\frac{\left[\left(5\Delta\right)^{2}+2\left(5\Delta\right)\omega_{L}\cos\theta_{m}
+\frac{3}{2}\omega_{L}\cos^{2}\theta_{m}\right]}{\left(5\Delta+\omega_{L}\cos\theta_{m}\right)^{3}},\nonumber\\
\nonumber
\end{eqnarray}
which vanishes, for example, in zero electric field. Generally, since the problem is nonlinear, the geometric phases
contain cross-coupling terms which depend on both the magnetic and electric parameters. This demonstrates that the total
geometric phase is not simply the additive result of the magnetic and electric contributions.

Before concluding this section, we make a comment about the situation where the magnetic and electric fields rotate at different
frequencies. In that case the usual Floquet theorem does not hold, and bichromatic Floquet theory needs to be used \cite{Chua2004}.
In that theory, the finite dimensional time-dependent matrix $H_{M}(t)$ is transformed into an \textit{infinite}-dimensional
time-independent dressed matrix. An analysis of the problem is being planned for the future.

\section{Conclusion}
\label{sec:Conc}
We have presented the geometric phase experienced by states in the ground state manifold of the OH molecule in the presence
of magnetic and electric fields varying harmonically in time. It is important to account for these phases systematically in
precision spectroscopy techniques such as the Ramsey method. Our calculations are exact within the description provided by
an effective Hamiltonian that neglects hyperfine structure, but has been shown to agree well with experimental data. We have
also supplied approximate expressions for the lowest order Berry phase as well as higher non-adiabatic corrections, where
appropriate. Our work shows specifically that the experimental parameters may be chosen to cancel the geometric phase acquired
by a single level or the relative geometric phase between two energy levels. These configurations should be useful to ongoing
work on precision spectroscopy and quantum computation with OH. Generally, we have found that the presence of an electric field
results in rich behavior of the geometric phase, offering possibilities for its coherent control. Our results may form the basis
for a future perturbation theoretic treatment that includes hyperfine structure, or investigations in a more general setting
where the fields may have multiple Fourier components in their time variation. To conclude, our work makes an advance towards
the accurate characterization of systematic shifts in the spectroscopy of ground state OH.

We would like to thank the JILA groups for earlier discussions about their experimental and theoretical work on OH. We also thank
S. Agarwal for useful discussions, and M.B. and S. M. are particularly grateful to Prof. V. Lindberg for making their collaboration
possible.

\end{document}